\documentclass[reprint,aps,pra,superscriptaddress,notitlepage]{revtex4-1}
\usepackage{amssymb,amsmath}
\usepackage{graphicx}
\usepackage{dcolumn}
\usepackage{multirow}
\usepackage{color}
\usepackage[english]{babel}
\usepackage[caption=false]{subfig}

\usepackage{color,soul}  % for highlighting text, as requested by npjQI
\sethlcolor{white}
\soulregister\cite7
\soulregister\ref7

\newcommand{\ket}[1]{\ensuremath{|{#1}\rangle}}

\begin{document}

\def\simlt{\mathrel{\lower .3ex \rlap{$\sim$}\raise .5ex \hbox{$<$}}}

\title{\textbf{\fontfamily{phv}\selectfont 
Extending the coherence of a quantum dot hybrid qubit}}
\author{Brandur Thorgrimsson}
\affiliation{Department of Physics, University of Wisconsin-Madison, Madison, WI 53706, USA}
\author{Dohun Kim}
\affiliation{Department of Physics and Astronomy, Seoul National University, Seoul 08826, South Korea}
\author{Yuan-Chi Yang}
\affiliation{Department of Physics, University of Wisconsin-Madison, Madison, WI 53706, USA}
\author{L. W. Smith}
\affiliation{Department of Physics, University of Wisconsin-Madison, Madison, WI 53706, USA}
%\thanks{Current address: Luke's Current Address}
\author{C. B. Simmons}
\affiliation{Department of Physics, University of Wisconsin-Madison, Madison, WI 53706, USA}
\author{Daniel R. Ward}
\thanks{Current address: Sandia National Laboratories, Albuquerque, NM 87185, USA.}
\affiliation{Department of Physics, University of Wisconsin-Madison, Madison, WI 53706, USA}
\author{Ryan H. Foote}
\affiliation{Department of Physics, University of Wisconsin-Madison, Madison, WI 53706, USA}
\author{J. Corrigan}
\affiliation{Department of Physics, University of Wisconsin-Madison, Madison, WI 53706, USA}
\author{D. E. Savage}
\affiliation{Department of Materials Science and Engineering, University of Wisconsin-Madison, Madison, WI 53706, USA}
\author{M. G. Lagally}
\affiliation{Department of Materials Science and Engineering, University of Wisconsin-Madison, Madison, WI 53706, USA}
\author{Mark Friesen}
\author{S. N. Coppersmith}
\author{M. A. Eriksson}
\affiliation{Department of Physics, University of Wisconsin-Madison, Madison, WI 53706, USA}

\begin{abstract}
Identifying and ameliorating dominant sources of decoherence are important steps in understanding and improving quantum systems. Here we show that the free induction decay time ($T_{2}^{*}$) and the Rabi decay rate ($\Gamma_{\mathrm{Rabi}}$) of the quantum dot hybrid qubit can be increased by more than an order of magnitude by appropriate tuning of the qubit parameters and operating points.  By operating in the spin-like regime of this qubit, and choosing parameters that increase the qubit's resilience to charge noise (which we show is presently the limiting noise source for this qubit), we achieve a Ramsey decay time $T_{2}^{*}$ of \hl{$177~\mathrm{ns}$} and a Rabi decay time $1/\Gamma_{\mathrm{Rabi}}$ exceeding $1~\mathrm{\mu s}$. We find that the slowest $\Gamma_{\mathrm{Rabi}}$ is limited by fluctuations in the Rabi frequency induced by charge noise and not by fluctuations in the qubit energy itself.
\end{abstract}

\maketitle

\section*{Introduction}

There has been much progress in the development of qubits in semiconductor quantum dots~\cite{Loss:1998p120}, making use of one~\cite{Koppens:2006p766,Simmons:2011p156804,Pla:2012p489,Veldhorst:2014p981,Kawakami:2014p666,Yoneda:2014p267601,Veldhorst:2015p410, Scarlino:2015p106802,House2016:p044016,Takeda:2016p1600694}, two~\cite{Levy:2002p1446,Petta:2005p2180, Foletti:2009p903, Petersson:2010p246804,Maune:2012p344,Wang:2013p046801,Shi:2013p075416,Wu:2014p11938,HarveyCollard:2015preprint}, and three quantum dots~\cite{DiVincenzo:2000p1642,Laird:2010p1985,Gaudreau:2011p54,Medford:2013p654,Eng:2015p1500214, Russ:2015p235411} to host qubits.  Charge noise is often the leading source of decoherence in semiconductor qubits~\cite{Dial:2013p146804}, and an advantage of using two or more quantum dots to host a single qubit is the ability to work at sweet spots, a technique pioneered in superconducting qubits~\cite{Vion:2002p886}, that make the qubit more resistant to charge noise~\hl{\cite{Taylor:2013p050502,Medford:2013p050501,Kim:2015p243,Fei:2015p205434,Reed:2016p110402,Martins:2016p116801,Shim:2016p121410,Nichol:2017p1}}.

In this work we focus on one such qubit, the quantum dot hybrid qubit (QDHQ)~\cite{Shi:2012p140503, Koh:2013p19695, Kim:2014p70, Ferraro:2014p1155, Mehl:2015preprint, DeMichielis:2015p065304, Cao:2016p086801, Wong:2016p035409, Chen:2017p035408}, which is formed from three electrons in a double quantum dot, and can be viewed as a hybrid of a spin qubit and a charge qubit.  Fast, full electrical control of the QDHQ was recently implemented experimentally using ac gating~\cite{Kim:2015p15004}, demonstrating a free induction decay (FID) time of 11~ns through operation in the spin-like operating region (see Fig.~1). While QDHQ gating times are fast, substantial further improvements in QDHQ coherence times are required to achieve the high-fidelity gating necessary for fault-tolerant operation~\cite{Fowler:2012p032324}.  True sweet spots, which are used to increase resistance to noise and thus increase coherence, are defined by a zero derivative of the qubit energy with respect to a parameter subject to noise. Sweet spots are usually found at specific points of zero extent in parameter space, so that non-infinitesimal noise amplitude temporarily moves a qubit off the sweet spot.  The spin-like regime of the QDHQ has no true sweet spot; however, it has a large and extended region of small $dE_Q/d\varepsilon$, where $E_Q$ is the qubit energy and $\varepsilon$ is the detuning between the two quantum dots.

Here we show that the spin-like operating regime for the QDHQ can be made resilient to charge noise by appropriate tuning of the internal parameters of the qubit.  By measuring $dE_Q/d\varepsilon$, we are able to identify dot tuning parameters that increase resiliency to charge noise.  These measurements show that the three-electron QDHQ can be tuned \emph{in-situ} in ways that have a predictable and understandable impact on the qubit coherence: the qubit dispersion can be tuned smoothly by varying device gate voltages, and we find that the dephasing rate is proportional to $dE_Q/d\varepsilon$, consistent with a charge noise dephasing mechanism. Reducing $dE_Q/d\varepsilon$ significantly enhances the coherence of the qubit.  We have achieved an increase the coherence times by more than an order of magnitude over previous work, decreasing the Rabi decay rate $\Gamma_\mathrm{Rabi}$ from 67.1~MHz to 0.98~MHz, and increasing the FID time $T_2^*$ to as long as \hl{177~ns}.  These parameters correspond to an infidelity contribution from pure dephasing of about 1\%.

\section*{Results}

Fig.~1 shows the energy levels of the QDHQ as a function of the detuning $\varepsilon$.
At negative detuning the energy difference between the $\ket{0}$ and $\ket{1}$ states is dominated by the Coulomb energy, while at large positive detunings, where both logical states have the same electron configuration (one electron on the left and two on the right), the energy difference is dominated by the single-particle splitting $E_{\mathrm{R}}$ between the lowest two valley-orbit states in the right dot.
Here the logical states are described by their spin configuration: $\ket{0}=\ket{\downarrow}\ket{S}$ and $\ket{1}=\sqrt{1/3}\ket{\downarrow}\ket{T_{0}}-\sqrt{2/3}\ket{\uparrow}\ket{T_{-}}$, where $\ket{\downarrow}$ and $\ket{\uparrow}$ represent the spin configuration of the single electron in the left quantum dot and $\ket{S}$, $\ket{T_{0}}$, and $\ket{T_{-}}$ represent the singlet (S) and triplet ($\mathrm{T}_{0},~\mathrm{T}_{-}$) spin configurations of the two electrons in the right quantum dot. 
The tunnel coupling $\Delta_{1(2)}$ describes the anticrossings between the right dot ground (first excited) state and left dot ground state.

\begin{figure}[h]
\includegraphics[width=0.5\textwidth]{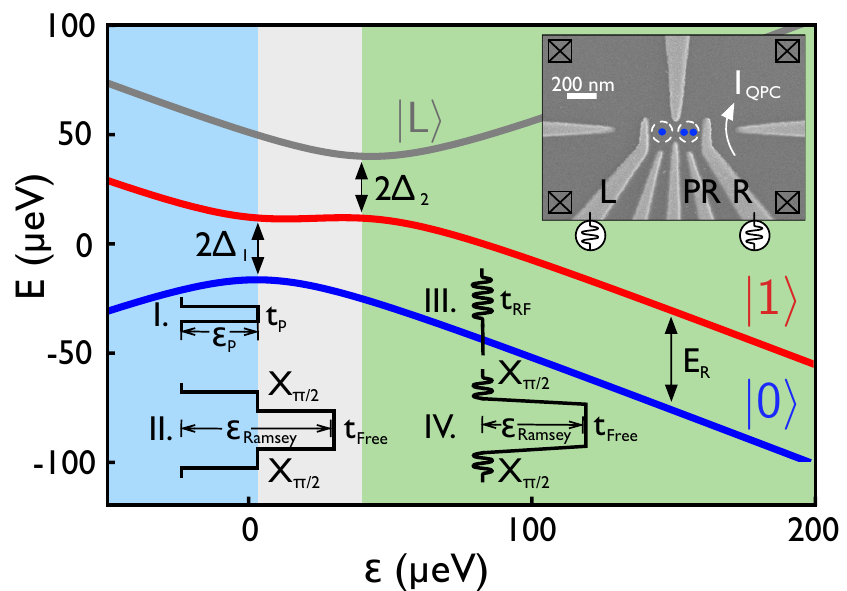}
\caption{
\textbf{Energy spectrum and pulse sequences for the quantum dot hybrid qubit (QDHQ).}
Main panel: Energy versus detuning of the qubit states $\vert 0 \rangle$ and $\vert 1 \rangle$ as well as a leakage state $\vert \mathrm{L}\rangle$.
The QDHQ Hamiltonian, described in Supplementary Section 1, is parameterized using two tunnel couplings $\Delta_{1(2)}$ between the ground state of the left dot and the ground (excited) state of the right dot, and the asymptotic energy splitting $E_{\mathrm{R}}$ between the ground and excited states of the right dot.
In the spin-like region (green, right), the logical states are differentiated by their spin configurations. The four pulse sequences used in this work are shown as functions of the detuning: the non-adiabatic Larmor (I) and Ramsey (II) sequences, and the microwave-pulsed Rabi (III) and Ramsey (IV) sequences. See Supplementary Section 4 for details.
\textbf{Inset}, SEM image of a device lithographically identical to the one used in the experiments; white dashed circles indicate the locations of the double dot. Voltage pulses are applied to gates L and R, and a quantum point contact (QPC) is used to measure the electron occupancy of the dots.
}
\end{figure}

\hl{Fig.~2a-g} shows results of FID measurements for four different values of the measured $dE_Q/d\varepsilon$, performed using the pulse sequence of diagram~IV of Fig.~1, in order to determine $\Gamma_{2}^{*}=1/T_2^*$.
For short times \hl{(panels a, c, e)}, Ramsey fringes are visible for all $dE_Q/d\varepsilon$; in contrast, for $t_\mathrm{Free}=22~\mathrm{ns}$, Ramsey fringes are attenuated in Fig.~2b (large $dE_\text{Q}/d\varepsilon$), yet are still clearly visible in Fig.~2f (small $dE_\text{Q}/d\varepsilon$).
As shown in Fig.~2\hl{g}, by tuning the qubit to achieve $dE_Q/d\varepsilon = 0.0025$, Ramsey fringes are still visible at $t_\mathrm{Free}=120$~ns, and at this tuning \hl{a Gaussian fit to the Ramsey fringe amplitude (shown in Fig.~2h) yields $T_2^* = 177\pm9~\mathrm{ns}$}.
Fits to the Ramsey fringe amplitude \hl{of the other three} detunings are shown in Fig.~2i, \hl{demonstrating a }strong correlation between small $dE_{\mathrm{Q}}/d\varepsilon$ and long $T_2^*$.
\hl{Although we have shown Gaussian fits in Fig.~2, consistent with quasistatic charge noise, we note that the FID decay also can be fit by an exponential decay, which would be consistent with noise that is dominated by only a few two-level fluctuators~\cite{Ithier:2005p134519}, and therefore we cannot distinguish between these two limiting cases (see supplemental material for fit parameters extracted from exponential decays).}

\begin{figure}[t!]
\includegraphics[width=0.5\textwidth]{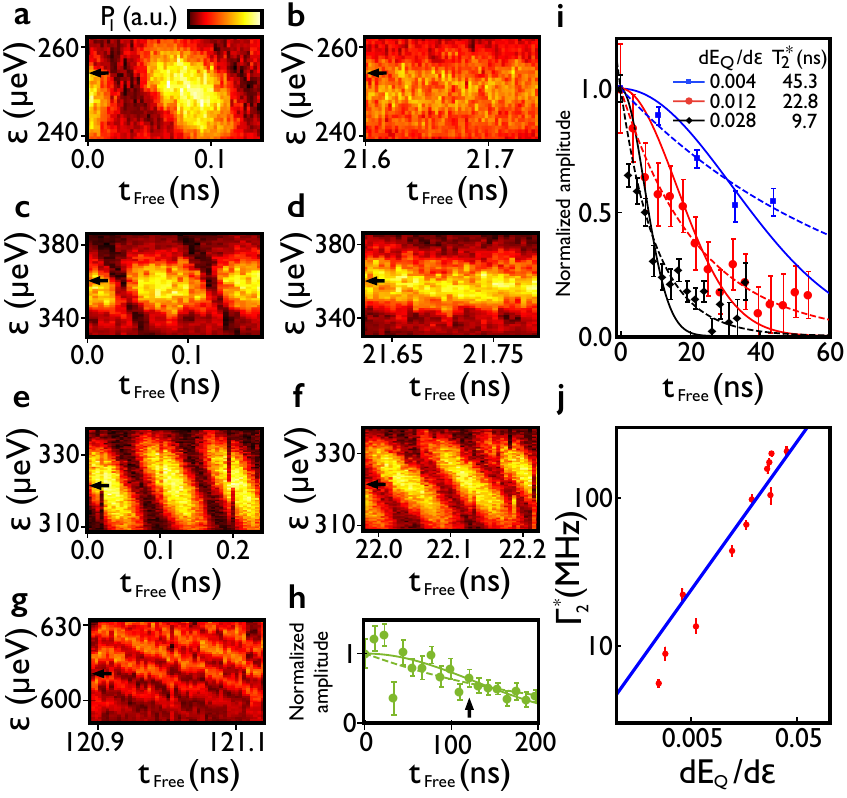}
\caption{
\textbf{Changing the dot tuning and $\varepsilon$ to achieve small $dE_{\mathrm{Q}}/d\varepsilon$ decreases the free induction decay (FID) rate by more than an order of magnitude.}
\textbf{a-\hl{g}} Plots showing the probability $P_1$ of being in state $\ket{1}$ after applying the Ramsey pulse sequence of diagram IV in Fig.~1, for qubit tunings characterized by different $dE_{\mathrm{Q}}/d\varepsilon$ values.
Two $t_\text{Free}$ time windows are shown for \hl{three tunings}, corresponding to $dE_{\mathrm{Q}}/d\varepsilon$=$0.028$ (\textbf{a,b}), $dE_{\mathrm{Q}}/d\varepsilon$=$0.012$ (\textbf{c,d}), $dE_{\mathrm{Q}}/d\varepsilon$=$0.0042$ (\textbf{e,f}), and \hl{a single time window is shown for} $dE_{\mathrm{Q}}/d\varepsilon$=$0.0025$ (\textbf{g}).
Comparing \textbf{b}, \textbf{d}, \textbf{f}, and \hl{\textbf{g}}, we see that the FID rate decreases as $dE_{\mathrm{Q}}/d\varepsilon$ decreases.
\textbf{\hl{h}, i}, Oscillation amplitudes as a function of $t_\text{Free}$, normalized by their value at ${t}_{\mathrm{Free}} = 0$ are obtained at the $\varepsilon$ values indicated by black arrows in \hl{\textbf{g} (\textbf{h}) and \textbf{a-f} (\textbf{i}); fits to both $\text{exp}(-(t_\text{Free}/T_2^*)^2)$ (values shown) and $\text{exp}(-t_\text{Free}/T_2^*)$ are plotted}.
\textbf{j}, $\Gamma_{2}^{*}$ vs. $dE_{\mathrm{Q}}/d\varepsilon$, obtained \hl{from a fit to $\text{exp}(-(t_\text{Free}/T_2^*)^2)$},  as in \textbf{i} \hl{(values extracted from a fit to $\text{exp}(-t_\text{Free}/T_2^*)$ can be found in Supplementary Section 6)}, for several different tunings and a range of $\varepsilon$.  The data are well fit to Eq.~(\ref{GammaEQ}) (blue line, \hl{$\sigma_{\varepsilon} = 4.39\pm 0.32~\mathrm{\mu eV}$}), providing evidence that $\Gamma_{2}^{*}$ is limited by charge noise.
}
\end{figure}

Fig.~2j shows $\Gamma_2^* = 1/T_{2}^{*}$ for a wide range of $dE_{\mathrm{Q}}/d\varepsilon$, \hl{demonstrating a significant improvement in coherence for reduced values of $dE_{\mathrm{Q}}/d\varepsilon$}.
For a gaussian distribution of quasistatic fluctuations of the detuning parameter, with a standard deviation of $\sigma_\varepsilon$, one expects that~\cite{Petersson:2010p246804,Dial:2013p146804}
\begin{equation}
	\Gamma_{2}^{*} = \rvert dE_{\mathrm{Q}}/d\varepsilon \lvert \sigma_{\varepsilon} / \sqrt{2}\hbar .
	\label{GammaEQ}
\end{equation}
In Fig.~2j, we observe such a linear relation between $\Gamma_2^*$ and $dE_\text{Q}/d\varepsilon$, with \hl{a} fitting constant \hl{$\sigma_{\varepsilon} = 4.39\pm 0.32~\mathrm{\mu eV}$.}

\begin{figure*}[t!]
\includegraphics[width=1\textwidth]{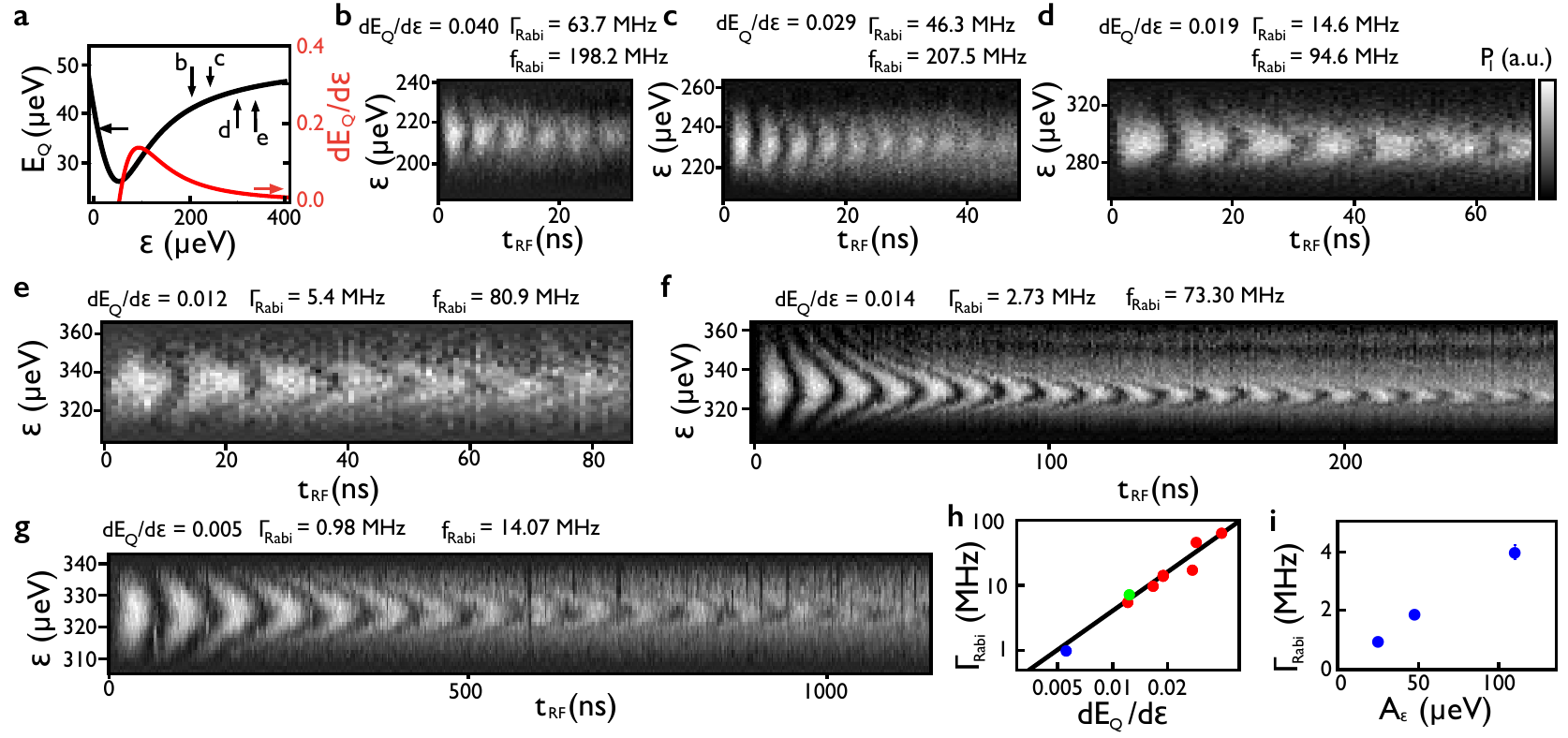}
\caption{
\textbf{Rabi decay rate is limited by charge noise and applied microwave power, $A_{\varepsilon}$.}
\textbf{a},
Plots of $E_{\mathrm{Q}}=hf_{\mathrm{Q}}$ (black) and $dE_{\mathrm{Q}}/d\varepsilon$ (red) versus $\varepsilon$ for the tuning used in panels \textbf{b-e}.  Here, $f_{\mathrm{Q}}$ is the qubit frequency, and the spectroscopy methods used are described in Supplementary Section 4. 
\textbf{b-e}, Rabi oscillations of the probability $P_1$ of being in state $\ket{1}$, all obtained at the same tuning but at different $\varepsilon$, ranging from $\sim210$ to $340 ~\mathrm{\mu eV}$.
$\Gamma_{\mathrm{Rabi}}$ clearly decreases as $dE_{\mathrm{Q}}/d\varepsilon$ decreases. \hl{The decrease in $f_{\mathrm{Rabi}}$ between panels \textbf{b-e} is caused by the decreased coupling to the left dot as $\varepsilon$ is increased (see Eq.~(\ref{frabiEQ})). $A_{\varepsilon}$ is nominally the same but changes slightly between panels \textbf{b-e} due to changes in $f_{\mathrm{Q}}$ as discussed in Supplementary Section 5.}
\textbf{f}, Rabi oscillations, taken at a different device tuning, demonstrating over 100 coherent $X_{\pi/2}$ rotations within a Rabi decay time. 
\textbf{g}, Rabi oscillations demonstrating a Rabi decay time longer than 1~$\mu s$, taken at a device tuning differing from those in \textbf{b-f}.
\textbf{h}, $\Gamma_{\mathrm{Rabi}}$, obtained by fitting to an exponential decay, plotted as a function of $dE_{\mathrm{Q}}/d\varepsilon$ for measurements at multiple tunings and operating points, with $A_{\varepsilon}$ within 10\% of $25~\mathrm{\mu eV}$. 
Here, the black line has slope~2, indicating that $\Gamma_{\mathrm{Rabi}}$ depends quadratically on $dE_{\mathrm{Q}}/d\varepsilon$, consistent with $\Gamma_{\mathrm{Rabi}}$ being limited by fluctuations of $f_{\mathrm{Q}}$~\cite{Jing:2014p022118,Ithier:2005p134519}. Here, the different tunings are labeled with different colors (red, green, and blue), as specified in Supplementary Section 1.  
\textbf{i}, Rabi oscillations taken at $\varepsilon=323$~$\mu$eV ($dE_{\mathrm{Q}}/d\varepsilon = 0.005$), as a function of the microwave amplitude $A_{\varepsilon}$, showing that the Rabi decay rate $\Gamma_{\mathrm{Rabi}}\propto A_{\varepsilon}$, consistent with $\Gamma_{\mathrm{Rabi}}$ being limited by fluctuations of $f_{\mathrm{Rabi}}$ for small values of $dE_{\mathrm{Q}}/d\varepsilon$.
}
\end{figure*}

We now turn to a discussion of the Rabi decay time, $1/\Gamma_{\mathrm{Rabi}}$, and its dependence on the qubit dispersion $dE_{\mathrm{Q}}/d\varepsilon$.  Fig.~3a shows both $E_{\mathrm{Q}}$ and $dE_{\mathrm{Q}}/d\varepsilon$ as a function of detuning, calculated using the measured tuning parameters for Figs.~3b-e (see Supplementary Section 1 and 4), showing the decrease in the slope $dE_{\mathrm{Q}}/d\varepsilon$ with increasing $\varepsilon$.
Figs.~3b-e show Rabi oscillation measurements, performed with a microwave burst of duration $t_{\mathrm{RF}}$ and acquired at the detunings labeled b-e in Fig.~3a, showing that with increasing $\varepsilon$ (and therefore decreasing $dE_{\mathrm{Q}}/d\varepsilon$) the Rabi decay rate $\Gamma_{\mathrm{Rabi}}$ decreases by more than an order of magnitude for the data reported here.

For quantum gates, the contribution to infidelity arising from qubit decoherence is minimized when the ratio of the gate duration to the Rabi decay time is minimized.  The data in Fig.~3f, acquired at a different dot tuning, shows that this ratio can be made small enough that an $X_{\pi/2}$ gate can be performed over 100 times within one Rabi decay time.
In the absence of any other nonideality in the experiment, this would limit the fidelity of an $X_{\pi/2}$ rotation on the Bloch sphere to $99.0\%$ and would represent a sevenfold improvement over previous results~\cite{Kim:2015p15004}.

It is also interesting to consider how long the Rabi decay time, $1/\Gamma_\mathrm{Rabi}$, itself can be. Fig.~3g shows Rabi oscillations acquired at a different dot tuning and a very small $dE_{\mathrm{Q}}/d\varepsilon = 0.005$.
Here, $\Gamma_{\mathrm{Rabi}} = 0.98~\mathrm{MHz}$, representing a decrease by more than a factor of 30 from previously reported Rabi decay rates~\cite{Kim:2015p15004}.

The decay of Rabi oscillations is caused by at least two different mechanisms~\cite{Yan:2013p2337}, both of which are observed in these experiments.
First, for relatively large values of $dE_{\mathrm{Q}}/d\varepsilon$, fluctuations in $E_{\mathrm{Q}}$ from charge noise dominate the decoherence. This is similar to FID measurements, with the important difference that the microwave drive effectively reduces the range of frequencies decohering the qubit. This results in Rabi decoherence rates $\Gamma_{\mathrm{Rabi}}$ that are slower than the FID rates $\Gamma_{2}^{*}$ at the same $dE_{\mathrm{Q}}/d\varepsilon$. For this mechanism, the Rabi decay is expected to be exponential and depend quadratically on $dE_{\mathrm{Q}}/d\varepsilon$~\cite{Jing:2014p022118,Ithier:2005p134519}.
Fig.~3h shows $\Gamma_{\mathrm{Rabi}}$ vs.\ $dE_{\mathrm{Q}}/d\varepsilon$ and a quadratic fit to the data; the data are well-described by this functional form, and decreasing $dE_{\mathrm{Q}}/d\varepsilon$ yields nearly two orders of magnitude decrease in $\Gamma_{\mathrm{Rabi}}$. 

Second, charge noise can also cause fluctuations in the rotation rate $f_{\mathrm{Rabi}}$ itself~\cite{Yan:2013p2337}, and as $dE_{\mathrm{Q}}/d\varepsilon$ becomes small, these fluctuations become the dominant source of decoherence.
This second decay process is expected to yield a decay rate proportional to the drive amplitude $A_{\varepsilon}$, and as shown in Fig.~3i, we observe this proportionality in the experiment for small $dE_{\mathrm{Q}}/d\varepsilon$. Thus, for small $dE_{\mathrm{Q}}/d\varepsilon$, fluctuations in $f_{\mathrm{Rabi}}$ dominate the Rabi decay rate. 
In contrast to the Rabi decay process discussed above, in which the applied microwave pulse narrows the frequency range of charge fluctuations contributing to the decay, charge fluctuations over a wide bandwidth are expected to contribute to this decay process.  This contribution can be seen by applying the rotating wave approximation to Eq.~(S1) in Supplementary Section 1, which yields an approximate form for $f_{\mathrm{Rabi}}$ that is valid at large detunings:
\begin{equation}
	f_{\mathrm{Rabi}} = \frac{\Delta_{1}\Delta_{2}}{2h\varepsilon(\varepsilon-E_{\mathrm{R}})}A_{\varepsilon}.
	\label{frabiEQ}
\end{equation}
$\sigma_{\varepsilon}$ can then be related to $\sigma_{\mathrm{Rabi}}$, the standard deviation of fluctuations in $f_{\mathrm{Rabi}}$, by
\begin{equation}
	\sigma_{\mathrm{Rabi}} = (df_{\mathrm{Rabi}}/d\varepsilon)\sigma_{\varepsilon}.
\end{equation}
We therefore expect the decay rate from this mechanism to be proportional to $dE_{\mathrm{Q}}/d\varepsilon$ rather than to the square of $dE_{\mathrm{Q}}/d\varepsilon$, explaining its dominance at small $dE_{\mathrm{Q}}/d\varepsilon$.

\section*{Discussion}

In this work we have shown that the internal parameters of the QDHQ can alter the qubit dispersion $dE_{\mathrm{Q}}/d\varepsilon$ over a wide range, resulting in large tunability in both the decoherence rates and the Rabi frequencies achievable.
The dominant dephasing mechanism for Rabi oscillations switches from fluctuations in the qubit energy $E_{\mathrm{Q}}$ to fluctuations in the Rabi frequency $f_{\mathrm{Rabi}}$ at the smallest values of $dE_{\mathrm{Q}}/d\varepsilon$.
By decreasing $dE_{\mathrm{Q}}/d\varepsilon$ we have reduced both the Rabi and the Ramsey decoherence rates, important metrics for achieving high-fidelity quantum gate operations, by more than an order of magnitude compared with previous work, demonstrating $\Gamma_\mathrm{Rabi}$ as small as 0.98~MHz and $T_2^* = 1/\Gamma_2^*$ as long as 127~ns.  These coherence times exhibit the utility of the extended near-sweet spot in the QDHQ for improving qubit performance in the presence of charge noise.
%\hl{Further advances in dot designs should substantially improve both coherence and speed of qubit operations by enhancing the tunability of interdot tunnel couplings and tunnel-lead couplings~\cite{Wang:2013p046801, Reed:2016p110402, Martins:2016p116801, Wong:2016p035409}.}

\section*{Methods}

The Si/SiGe device is operated in a region where magnetospectroscopy measurements~\cite{Simmons:2011p156804,Shi:2011p233108} have indicated that the valence electron occupation of the double dot is (1,2) for the qubit states studied here.
Manipulation pulse sequences were generated using Tektronix 70001A arbitrary waveform generators and added to DC gate voltages on gates L and R using bias tees (PSPL5546).
Because of the frequency dependent attenuation of the bias tees, corrections were made to the applied pulses during the adiabatic detuning pulses, as described in Supplementary Section 5.
The qubit states were mapped to the (1,1) and (1,2) charge occupation states as described in ref.~\cite{Kim:2015p15004}.
A description of the methods used to measure the qubit dispersion and lever arm can be found in Supplementary Section 4.\\

Correspondence and requests for materials should be addressed to Mark A. Eriksson (maeriksson\emph{@}wisc.edu)

\section*{Acknowledgements}
This work was supported in part by ARO (W911NF-12-0607, W911NF-08-1-0482), NSF (DMR-1206915, PHY-1104660, DGE-1256259), and the Vannevar Bush Faculty Fellowship program sponsored by the Basic Research Office of the Assistant Secretary of Defense for Research and Engineering and funded by the Office of Naval Research through grant N00014-15-1-0029. Development and maintenance of the growth facilities used for fabricating samples is supported by DOE (DE-FG02-03ER46028). This research utilized NSF-supported shared facilities at the University of Wisconsin-Madison.

\section*{Supplemental material for `Extending the coherence of a quantum dot hybrid qubit'}

These supplemental materials present additional details of the experimental methods used.

\section*{S1. Hamiltonian of the quantum dot hybrid qubit}
The Hamiltonian of the quantum dot hybrid qubit (QDHQ) can be written as
\begin{equation}
\begin{pmatrix}
	\varepsilon/2 & \Delta_1 & -\Delta_2\\
	\Delta_1 & -\varepsilon/2 & 0\\
	-\Delta_2 & 0 & -\varepsilon/2 + E_{\mathrm{R}}
\end{pmatrix} .
\label{Hamiltonian}
\end{equation}
If the tunnel couplings are assumed to be independent of $\varepsilon$, then $\Delta_1$, $\Delta_2$ and $E_{\mathrm{R}}$ can be determined by fitting the spectrum obtained from Eq.~(\ref{Hamiltonian}) to the measured energy difference between the two lowest states of the QDHQ, $E_\text{Q}(\varepsilon)$. 
Table~\ref{tab1} shows the values of $\Delta_{1}$, $\Delta_{2}$ and $E_{\mathrm{R}}$ obtained in this way for the 7 different dot tunings used in this experiment, as described below.
The data in Figs.~2(a)-(b) is obtained at tuning 4, the data in Figs.~2(c)-(d) is obtained at tuning 1 while the data in Figs.~2(e)-(h) is obtained at tuning 3. For the data in Figs.~2(a)-(f) we used microwave pulsed Ramsey sequences with $\varepsilon_\text{Ramsey}=0~\mathrm{\mu eV}$ while in Figs.~2(e)-(h) $\varepsilon_\text{Ramsey}=288.6~\mathrm{\mu eV}$. The data in Figs.~3(b)-(e) are obtained at tuning 1, The data in Fig.~3(f) is obtained at tuning 7 while the data in Figs.~3(g) and (i) are acquired at tuning 3. In Fig.~3(h) red points are taken at tuning 1, the green point at tuning 2 and the blue point at tuning 3, here $A_{\varepsilon}$ was kept within 10\% of $25~\mathrm{\mu eV}$, as described in Sec.~S5. For both Rabi and Ramsey measurements we use the measured energy spectrum to determine the resonant detuning value, $\varepsilon$, corresponding to the microwave frequency, $f_{\mathrm{\mu w}}$, used in the experiments.
%The data in Figs.~4(a)-(b) is obtained at tuning 4 while the data in Figs.~4(c)-(d) is obtained at tuning 1.
\begin{center}
\begin{table}[h]
\begin{tabular}{c|c|c|c|c}
	Tuning & Method & $\Delta_{1}~(\mu\mathrm{eV})$  & $\Delta_{2}~(\mu\mathrm{eV})$& $E_{\mathrm{R}}~(\mu\mathrm{eV})$ \\
	\hline
	1 & I &	$18.1\pm1.1$ & $46.7\pm3.1$ & $51.7$ \\%020916-030616
	2 & IV &	$18.0\pm5.9$ & $49.0\pm11.9$ & $53.6\pm9.4$ \\%032216-032316
	3 & IV &    $16.1\pm1.7$ & $35.5\pm4.2$	& $50.6\pm3.7$ \\%042016-050416
	4 & I &      $16.9\pm2.4$ & $41.1\pm3.4$ & $49.6\pm2.7$ \\%081415-090115
	5 & II &    $21.1\pm2.6$ & $67.3\pm5.3$ & $58.6\pm5.3$\\ %052015-062815
	6 & IV &	$15.5\pm1.3$ & $33.4\pm3.7$ & $50.6\pm2.7$\\ %092816	USED TO DETERMINE ALPHA VALUE
	7 & IV & $17.3\pm0.8$ & $50.2\pm1.3$ & $52.0$ %100716-101116 BEST RABI 

\end{tabular}
\caption{Values of $\Delta_{1}$, $\Delta_{2}$ and ${E}_{\mathrm{R}}$ for each of the tunings used in this experiment. `Method' refers to the Larmor or Ramsey pulse sequences indicated schematically in Fig.~1 of the main text.}
\label{tab1}
\end{table}
\end{center}

\section*{S2. Experimental setup}
The experiments were performed in a device with a gate design identical to the one shown in Fig.~1 of the main text, which was placed in a dilution refrigerator with a base temperature $\leq 30~\mathrm{mK}$. The electron temperature is estimated to be $143\pm10~\mathrm{mK}$.~\cite{Simmons:2011p156804}.
The device was operated near the (2,1)-(1,2) charge transition.
Here, $(n,m)$ refers to a charge occupation of $n$ electrons in the left dot and $m$ electrons  in the right dot, modulo a possible closed shell of electrons in one or both dots.
All pulse sequences were generated using a Tektronix AWG70001A arbitrary waveform generator (AWG).
Manipulations using non-adiabatic pulse sequences were performed by applying pulses while detuned to the charge-like (CL) region, highlighted blue (left part) in Fig.~1, where we initialize to a (2,1) charge occupation. 
In the charge-like region the two logical states have different charge configurations and are well described by $\ket{0}_\text{CL}=\ket{\text{L}}$ and $\ket{1}_\text{CL}=\ket{\text{R}}$, where state L (R) corresponds to two electrons in the left (right) quantum dot and one electron in the right (left) quantum dot. This enables readout by projecting the $\ket{0}_\text{CL}$ state onto (2,1) and the $\ket{1}_\text{CL}$ state onto (1,2).
Manipulations using microwave pulse sequences were performed by applying microwaves while detuned to the spin-like (SL) region, highlighted green (right part) in Fig.~1, where the initial charge occupation is (1,2).
To perform readout, spin-to-charge conversion is implemented near the (1,1) to (1,2) charge transition line, projecting the $\ket{1}_\text{SL}$ state onto a (1,1) charge configuration and the $\ket{0}_\text{SL}$ onto a (1,2) charge configuration. We measure the qubit charge configuration using the methods described in~Ref.~\cite{Kim:2015p15004}.

\section*{S3. Calibration of pulse amplitudes to gate biases}

\begin{figure}[t]
%\vspace{-10pt}
\includegraphics[width=0.5\textwidth]{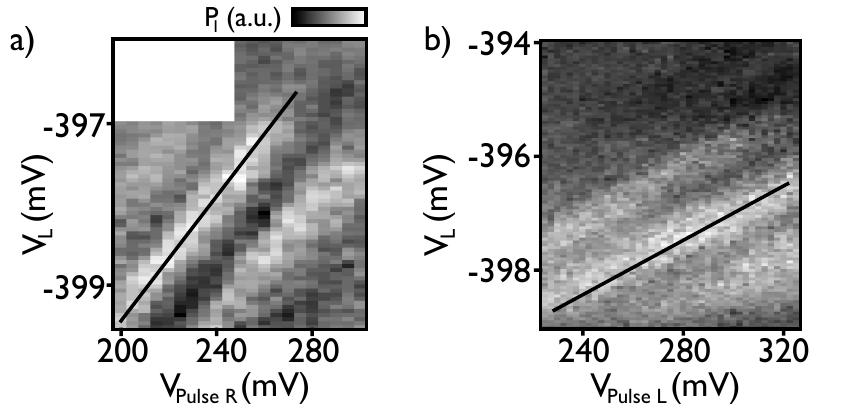}
\caption{ \textbf{Calibrating the amplitudes of high-frequency pulses on the L and R gates to dc voltage changes on the L gate.}
\textbf{a}, The probability of being in the qubit state $\ket{1}$ after applying a 1~ns Larmor pulse (shown in diagram I of Fig.~1), resulting in Larmor oscillations as a function of the ac pulse amplitude on the R gate, ${V}_{\mathrm{Pulse~R}}$, and the dc voltage bias applied to gate L, ${V}_{\mathrm{L}}$.  
The black line identifies one of the Larmor peaks.  In this case, a change of the Larmor pulse amplitude of $\delta{V}_{\mathrm{Pulse~R}} = 28.8~\mathrm{mV}$, corresponds to a shift in the dc bias on the L gate by $\delta{V}_{\mathrm{L}} = 1.0~\mathrm{mV}$.
\textbf{b}, Analogous result when the Larmor pulse is applied to gate L.  In this case, a change of the Larmor pulse amplitude by $\delta{V}_{\mathrm{Pulse~L}} = 47.4~\mathrm{mV}$, corresponds to a shift in the dc bias on the L gate by $\delta{V}_{\mathrm{L}} =1.0~\mathrm{mV}$.
}
\label{pulse_cal} 
\vspace{-8pt}
\end{figure}

The L and R gates can be used to apply high-frequency pulses, including non-adiabatic and microwave-driven pulses.
(For simplicity here, we refer to both of these as ac pulses.)
The electrostatic coupling between the gates and the dots is the same for any type of pulse.
However, ac pulses are intentionally attenuated in our control circuitry, which effectively suppresses the ac lever arm.
We can calibrate changes in the amplitude of applied pulses to changes in applied bias on the L gate by applying a 1~ns Larmor pulse to the L(R) gate and sweeping the bias of the L gate over the (2,1)-(1,2) charge polarization line. As the bias is swept, Larmor oscillations cause oscillations in the QPC current. Changing the amplitude of the pulse from $V_{\mathrm{pulse~L(R)}} =  225~(200)~\mathrm{mV}$ to $325~(300)~\mathrm{mV}$ causes a shift in the position of the Larmor oscillation with respect to bias on the L gate, as shown in Fig.~\ref{pulse_cal}. For a $\delta{V}_{\mathrm{pulse~L(R)}} = 47.4~(28.8)~\mathrm{mV}$ change in pulse amplitude applied to the L(R) gate the shift was measured to be a $\delta{V}_{\mathrm{L}} = 1.0~\mathrm{mV}$ bias change on the L gate, yielding an effective ratio of pulse amplitudes to gate bias of
\begin{equation}
%	\alpha_{\mathrm{\varepsilon,L(R)}}^{\mathrm{ac}} = \alpha_{\mathrm{\varepsilon,L}}\frac{\delta\mathrm{V}_{\mathrm{L}}}{\delta\mathrm{V}_{\mathrm{pulse~L(R)}}} = 0.30~(0.49)~\mathrm{\mu eV/mV}.
	\frac{\delta\mathrm{V}_{\mathrm{L}}}{\delta\mathrm{V}_{\mathrm{pulse~L(R)}}} = 0.021~(0.035).
\end{equation}

\begin{figure*}[t]
\includegraphics[width=1.0\textwidth]{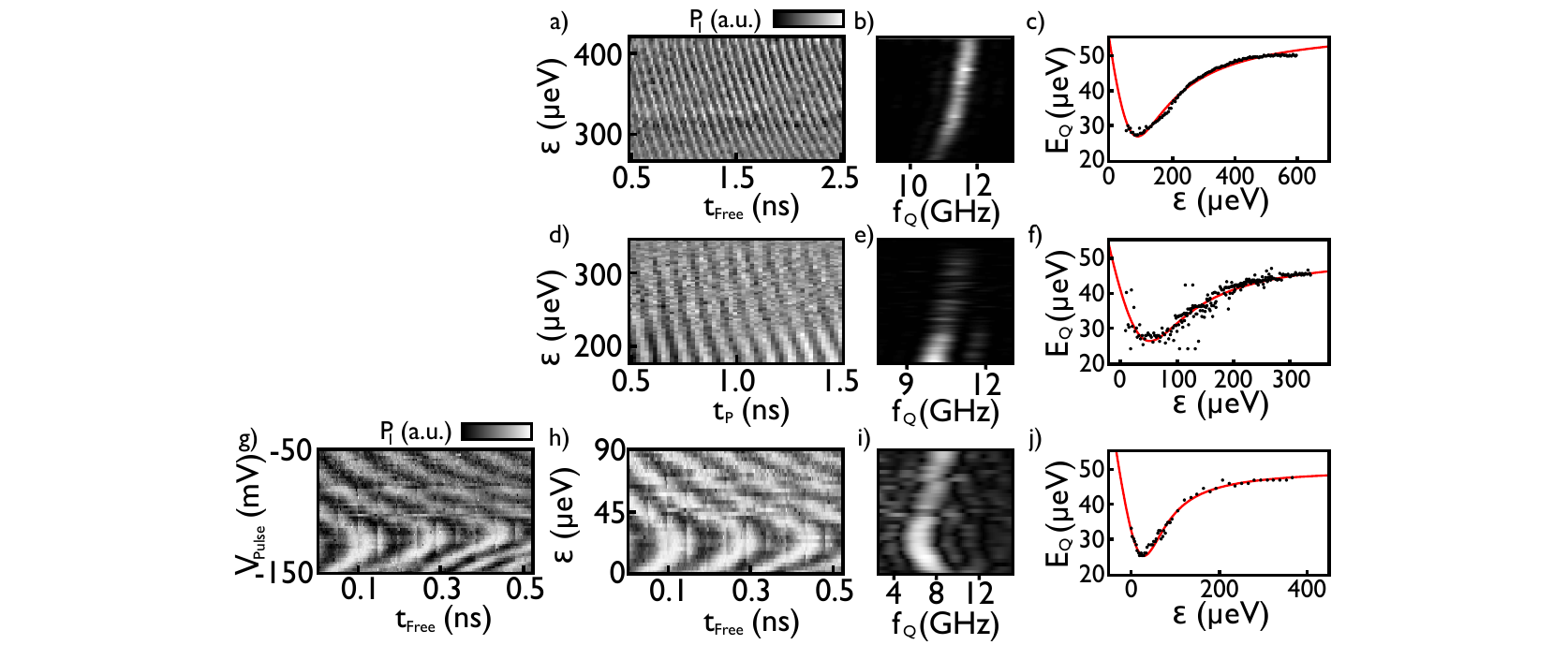}
\caption{
\textbf{Qubit spectroscopy measurements, using three different methods. }
\textbf{a-c}, The non-adiabatic Ramsey pulse method, shown schematically in diagram~II of Fig.~1 in the main text.
\textbf{a}, The gray-scale image shows the probability of being in qubit state $\ket{1}$ as a function the wait time, $t_\text{Free}$, and the detuning, $\varepsilon = \varepsilon_{0}+\varepsilon_{\mathrm{Ramsey}}$, where free induction occurs.
Here, the base detuning, $\varepsilon_0$, is swept from $-290$ to $-175$~$\mu$eV, while the pulse amplitude $\varepsilon_\text{Ramsey}=505$~$\mu$eV remains constant for the whole scan.
\textbf{b}, A time Fourier transform of the Ramsey oscillations in (\textbf{a}); the qubit frequency, $f_{\mathrm{Q}}$, is identified as the location of the peak in the Fourier transform as a function of $\varepsilon$.
\textbf{c}, $E_{\mathrm{Q}} = hf_{\mathrm{Q}}$ vs.\ $\varepsilon$, obtained by combining different Ramsey fringe measurements at tuning 5, with $\varepsilon_\text{Ramsey}$ in the range of 100-640~$\mu$eV.
The red curve shows a fit to the QDHQ energy splitting, modelled in Eq.~(\ref{Hamiltonian}), assuming constant tunnel couplings; the resulting fitting parameters for tuning~5 are listed in Table~\ref{tab1}.
\textbf{d-f}, The non-adiabatic Larmor pulse method, shown schematically in diagram~I of Fig.~1.
\textbf{d}, The gray-scale image shows the probability of being in qubit state $\ket{1}$ as a function wait time, $t_\text{p}$, and the detuning, $\varepsilon = \varepsilon_{0}+\varepsilon_{\mathrm{p}}$, where the Larmor rotation is performed.
In this case, the base detuning $\varepsilon_{0}$ is swept from $-213$ to $-91~\mathrm{\mu eV}$, while $\varepsilon_{\mathrm{p}}= 350~\mathrm{\mu eV}$ is held constant. 
\textbf{e}, A time Fourier transform of the data in (\textbf{d}), with the qubit frequency again identified as the location of the peak.
\textbf{f}, $E_{\mathrm{Q}}$ vs. $\varepsilon$ obtained by combining different Larmor fringe measurements at tuning~1, corresponding to $\varepsilon_{\mathrm{p}}$ in the range of 175-350~$\mathrm{\mu eV}$.
The red curve shows a fit to the model of Eq.~(\ref{Hamiltonian}), with constant tunnel couplings, yielding the results for tuning~1 listed in Table~\ref{tab1}.
\textbf{g-j}, The microwave-driven Ramsey pulse method, shown schematically in diagram~IV of Fig.~1.
\textbf{g}, The gray-scale image shows the probability of being in qubit state $\ket{1}$ as a function wait time, $t_\text{Free}$, and the uncalibrated pulse amplitude of the adiabatic pulse, $\mathrm{V}_{\mathrm{Pulse}}$, where the free induction occurs.
Here, the base detuning is kept constant at $\varepsilon_{0}=210~\mathrm{\mu eV}$, while $\mathrm{V}_{\mathrm{Pulse}}$ is swept from $-150$ to $-100~\mathrm{mV}$. 
\textbf{h}, The same data as (\textbf{g}), replotted as a function of $\varepsilon = \varepsilon_{0} + \varepsilon_{\mathrm{Ramsey}}$.  In this case, $\mathrm{V}_{\mathrm{Pulse}}$ has been scaled to take into account the frequency-dependent attenuation that occurs in the circuit between the arbitrary waveform generator and the sample, which distorts the pulse shape at short times, (see Sec.~S5.) and then scaled to $\varepsilon_{\mathrm{Ramsey}}$ using $\alpha_{\mathrm{\varepsilon,R}}^{\mathrm{ac}}$ (see Sec.~S4).
\textbf{i}, A time Fourier transform of the data in (\textbf{h}), with the qubit frequency identified as the location of the peak.
\textbf{j}, $E_{\mathrm{Q}}$ vs.\ $\varepsilon$ obtained by combining different microwave-driven Ramsey fringe measurements at tuning~3.
The red curve shows a fit to the model of Eq.~(\ref{Hamiltonian}), with constant tunnel couplings, yielding the results for tuning~3 listed in Table~\ref{tab1}. }
\label{spectrum}
\vspace{-8pt}
\end{figure*}

\section*{S4. Measuring the qubit energy spectrum and determining the lever arms}
The energy difference between the two qubit states was measured as a function of detuning using three different techniques: non-adiabatic Larmor and Ramsey pulse sequences, corresponding to diagrams I and II in Fig.~1 of the main text, and a microwave-pulsed Ramsey sequence corresponding to diagram IV.

For the microwave pulsed Ramsey sequence, initialization and readout are performed in the spin-like regime at the base detuning $\varepsilon_0$, using state-dependent tunneling between the charge configurations $(1,1)$ and $(1,2)$, as described in Sec.~S2 and Ref.~\cite{Kim:2015p15004}.  \hl{First we perform an $X_{\pi/2}$ rotation} at $\varepsilon=\varepsilon_0$. The microwave driving frequency is set to the qubit resonant frequency $f_{\mathrm{\mu w}} = f_{\mathrm{Q}}(\varepsilon_{0}) = E_{\mathrm{Q}}(\varepsilon_{0})/h$, \hl{and the pulse duration ranges from 1.2 to 17.7~ns, depending on the tuning of the device}.  \hl{These microwave pulses are turned on and off smoothly over 200~ps, time periods that are included in the total pulse duration.}
\hl{After the first microwave burst} we adiabatically, \hl{over a 400~ps time period, change} the detuning to $\varepsilon=\varepsilon_{0}+\varepsilon_{\mathrm{Ramsey}}$ \hl{as shown in diagram IV in Fig.~1 of the main text}, so that the probabilities of the logical states are unaffected.
Phase is accumulated between the logical states during the wait time $t_\text{Free}$, \hl{after which the detuning is adiabatically changed back to $\varepsilon=\varepsilon_0$, again over a 400~ps time period, before performing a second microwave-driven $X_{\pi/2}$ rotation}.
This procedure yields Ramsey oscillations as a function of the wait time, $t_\text{Free}$, with a frequency given by $E_{\mathrm{Q}}(\varepsilon_{0}+\varepsilon_{\mathrm{Ramsey}})/h$.
Hence, by measuring the Ramsey oscillation frequency, we can determine the qubit energy, $E_\text{Q}$.
By varying $\varepsilon_{0}$ or $\varepsilon_{\mathrm{Ramsey}}$, the energy spectrum can be mapped out over a wide range of detuning values.
%\textcolor{red}{The microwave pulses were adiabatically turned on an off over a 200~ps time period. The length of the adiabatic ramp was 400 ps.}
The microwave method was employed at tunings 2, 3 [Figs.~\ref{spectrum}(g)-(j)], 6 and 7 to obtain the corresponding energy spectra. Fitting the data to Eq.~(\ref{Hamiltonian}) yields the results shown in Table~\ref{tab1}.
The lever arm $\alpha_{\mathrm{\varepsilon,L}}$ was determined by replacing $\varepsilon$ in Eq.~(\ref{Hamiltonian}) with $\alpha_{\mathrm{\varepsilon,L}}(V_{L}-V_{L,0})$, where $V_{L,0}$ is the $V_{L}$ bias corresponding to $\varepsilon = 0~\mathrm{\mu eV}$, and fitting for $\alpha_{\mathrm{\varepsilon,L}}$. The quality of this fit increases as more points are acquired at negative $\varepsilon$ values where $dE_{Q}/d\varepsilon \rightarrow -1$. The greatest number of points in this regime were acquired at tuning 6, where a fit yielded
\begin{equation}
	 \alpha_{\mathrm{\varepsilon,L}} = 41.36~\mathrm{\mu eV/mV}.
\end{equation}
We then calculate the lever arm between detuning and ac pulses:
\begin{equation}
	\alpha_{\mathrm{\varepsilon,L(R)}}^{\mathrm{ac}} = \alpha_{\mathrm{\varepsilon,L}}\frac{\delta\mathrm{V}_{\mathrm{L}}}{\delta\mathrm{V}_{\mathrm{pulse~L(R)}}} = 0.87~(1.46)~\mathrm{\mu eV/mV}.
\end{equation}
The microwave driving method was also used to obtain the data in Figs.~2 and 3 of the main text.
When using the method to determine the FID rate, $\Gamma_{2}^{*}$, it is inconvenient to increase $t_\text{Free}$ to large times, $>T_{2}^{*}$, because the decoherence rate in the spin-like region is very slow compared to the qubit frequency, and measuring for both long times and with short time steps is prohibitively time-consuming. 
Instead, we obtain a series of widely separated time windows, with good resolution in each window. These windows range in size from 120 to $400~\mathrm{ps}$, encompassing 1.5-6 oscillations, and the separation between the windows is $0.6-22.0~\mathrm{ns}$, getting larger as $\Gamma_2^*$ decreases. A given $\Gamma_2^*$ measurement includes at least 5 such windows.

For the non-adiabatic Ramsey pulse sequence shown schematically in diagram II of Fig.~1, the qubit begins in its ground state at a base detuning value of $\varepsilon_{0} \ll 0$ in the charge-like region.
We then pulse the detuning to the avoided crossing at $\varepsilon=0$, where the energy eigenstates are $\ket{0}_{\varepsilon=0}=(\ket{0}_\text{CL}-\ket{1}_\text{CL})/\sqrt{2}$ and $\ket{1}_{\varepsilon=0}=(\ket{0}_\text{CL}+\ket{1}_\text{CL})/\sqrt{2}$.
Waiting at this position for a time ${t}_{\mathrm{p}}$ yields an $\text{X}_{5\pi/2}$ rotation in the original basis, such that the qubit is in state $\ket{-Y}_\text{CL}=(\ket{0}_\text{CL}-i\ket{1}_\text{CL})/\sqrt{2}$. 
We note that a $5\pi/2$ rotation was used here, instead of a $\pi/2$ rotation, because the $20~\mathrm{ps}$ time resolution of our AWG allows us to better match the period of a $5\pi/2$ rotation than a $\pi/2$ rotation.
We then pulse to another detuning value, $\varepsilon_{0}+\varepsilon_{\mathrm{Ramsey}}$, and wait for a time ${t}_{\mathrm{Free}}$. 
This results in a phase accumulation of $\psi = e^{-i{t}_{\mathrm{Free}}E_{\mathrm{Q}}(\varepsilon_{0}+\varepsilon_{\mathrm{Ramsey}})/\hbar}$ between the logical states, $\ket{0}_{\varepsilon_{0}+\varepsilon_{\mathrm{Ramsey}}}$ and $\ket{1}_{\varepsilon_{0}+\varepsilon_{\mathrm{Ramsey}}}$.
%In the original basis, this corresponds to a rotation of angle $\psi$ about an axis in the $x$-$z$ plane whose orientation depends on $\varepsilon_\text{Ramsey}$, but is unimportant for this measurement.
We then pulse back to the avoided crossing, $\varepsilon = 0$, and perform another $\text{X}_{5\pi/2}$ rotation. Finally we pulse back to the base detuning, $\varepsilon_{0}$, and readout the qubit by measuring its charge configuration. 
Similar to the microwave pulsed Ramsey method, the microwave-driven sequence yields oscillations as a function of the wait time $t_\text{Free}$, with frequency $E_{\mathrm{Q}}(\varepsilon_{0}+\varepsilon_{\mathrm{Ramsey}})/h$.
This method was used in Figs.~\ref{spectrum}(a)-(c) to determine the qubit energy spectrum at tuning 5. 
Fitting the results to Eq.~(\ref{Hamiltonian}) yields the fitting parameters shown in Table~\ref{tab1}.

Finally, the non-adiabatic Larmor pulse sequence is shown schematically in diagram I of Fig.~1. It involves pulsing the detuning from its base value, $\varepsilon_{0} \ll~0$, to the position $\varepsilon_{0} + \varepsilon_{\mathrm{p}}$.
After waiting for a time ${t}_{\mathrm{p}}$, the logical states of the qubit, $\ket{0}_{\varepsilon_{0}+\varepsilon_{\mathrm{p}}}$ and $\ket{1}_{\varepsilon_{0}+\varepsilon_{\mathrm{p}}}$, accumulate a phase difference of $\phi = e^{-i{t}_{\mathrm{p}}E_{\mathrm{Q}}(\varepsilon_{0}+\varepsilon_{\mathrm{p}})/\hbar}$, where $E_{\mathrm{Q}}(\varepsilon_{0}+\varepsilon_{\mathrm{p}})$ is the qubit energy splitting at $\varepsilon=\varepsilon_{0}+\varepsilon_{\mathrm{p}}$.
Readout is then performed by pulsing back to $\varepsilon_{0}$ and measuring the charge configuration. 
By varying $\varepsilon_{0}$ and $\varepsilon_{\mathrm{p}}$, the qubit energy spectrum can again be measured over a wide range of detuning values.
This method was employed at tunings 1 [Figs.~\ref{spectrum}(d)-(f)] and 4 to obtain the corresponding energy spectra.  Fitting these data to Eq.~(\ref{Hamiltonian}) yields the results shown in Table~\ref{tab1}.

\begin{figure}[t]
%\vspace{-10pt}
\includegraphics[width=0.47\textwidth]{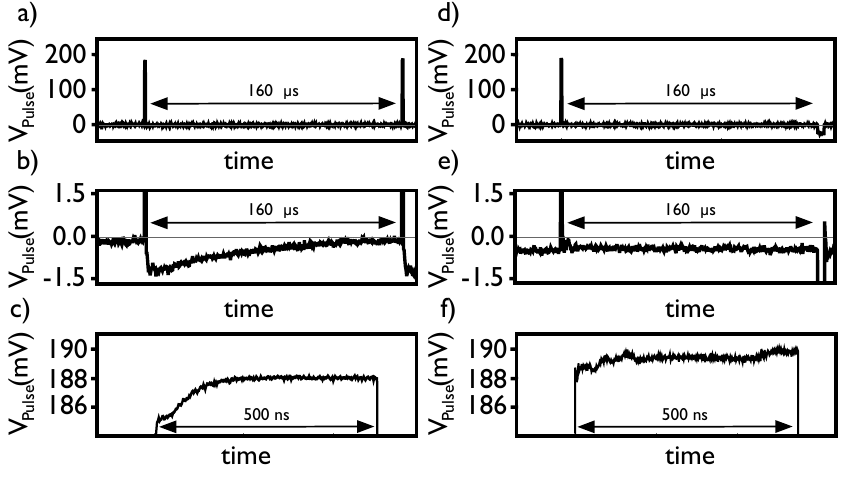}
\caption{
\textbf{Scope traces of an adiabatic pulse before and after pulse corrections.}
\textbf{a}, A 500~ns adiabatic pulse with a 0.5~ns in and out adiabatic ramps after passing through a bias-tee. 
The pulse is repeated every $160~\mathrm{\mu s}$.
\textbf{b}, Zoomed-in image of the bottom of the pulse in \textbf{(a)} showing distortions due to frequency-dependent attenuation.
\textbf{c}, Zoomed-in image of the top of the pulse in \textbf{(a)}, showing distortions due to frequency-dependent attenuation. 
The behavior of both the top and bottom parts was fitted by subtracting multiple exponential decays of different amplitudes and time constants from the desired pulse shape, Eq.~(\ref{Corrections}).
\textbf{d}, A 500~ns adiabatic pulse with a 0.5~ns in and out adiabatic ramps and pulse corrections applied that compensate for the frequency-dependent attenuation. 
Here, instead of repeating the pulse every $160~\mathrm{\mu s}$, we interleave a pulse with lower opposite amplitude but equal area between measurement pulses to reduce the dc offset.The pulse corrections, in this case, have the same time constants and relative amplitudes obtained by fitting the uncorrected pulse.
\textbf{e}, Zoomed-in image of the bottom of the corrected pulse shown in \textbf{(d)}.
\textbf{f}, Zoomed-in image of the top of the corrected pulse shown in \textbf{(d)}.
}
\label{Pulse_corr}
\vspace{-8pt}
\end{figure}

\section*{S5. Methods for correcting distortions in high frequency pulse sequences}
An arbitrary waveform generator (AWG) allows for the creation and application of complex pulse sequences.
Frequency-dependent attenuation caused by non-ideal RF-components of the experimental apparatus can cause distortions in the pulse shape before reaching the sample. 
Such is the case with the long adiabatic pulses used to measure the FID rate at the highest measured detunings. 
Here, the bandwidth between the AWG and the sample is limited by a bias-tee (PSPL5546) located in the coldest stage of the refrigerator ($\le30~\mathrm{mK}$) with a bandwidth from 3.5~KHz to 7~GHz. 
To compensate for this bandwidth limitation we applied a long adiabatic pulse through an identical bias-tee at room temperature and measured its shape using an oscilloscope (Tektronix DPO72504D), and then applied corrections to the generated pulse until the desired pulse shape was displayed on the oscilloscope.
Figs.~\ref{Pulse_corr}(a)-(c) show the shape of an uncorrected pulse after passing through a bias-tee. 
To compensate for the consistent dc offset seen in Fig.~\ref{Pulse_corr}(b) we interleave a second pulse that has the same area (pulse length $\times$ pulse amplitude) and period as the measurement pulse but opposite amplitude. 
We also find that adding an offset ($A_{b,\mathrm{off}}$) after the measurement pulse but not after the interleaved pulse further reduces the dc offset.
To compensate for the shape distortions we fitted 4 different exponential decays with different amplitudes $A_{t(b),i}$ and time constants $\tau_{t(b),i}$ to the difference of the measured pulse top (bottom) to that of the target pulse. 
All these corrections can be specified using 17 different parameters that are listed in Table.~\ref{pulse_parameters}.

\begin{center}
\begin{table}[ht]
\begin{tabular}{c|c|c|c}
	Correction (i) & $\tau_{t(b),i}~(\mathrm{ns})$ & $A_{t(b),i}$ & Scaling of $A_{t(b),i}~(\mathrm{mV})$ \\
	\hline
	t,1 & 1.25 & 19.5 & Amplitude\\
	t,2 & 5.00 & -2.0 & Amplitude\\
	t,3 & 30.0 & -0.8 & Amplitude\\
	t,4 & 120 & 4.0 & Amplitude\\
	b,1 & 1.25 & -17.5 & Area\\
	b,2 & 5.00 & 2.0 & Area\\
	b,3 & 80.0 & -3.0 & Area\\
	b,4 & 1560 & 1.5 & Area\\
	off & - & 2.5 & Area
\end{tabular}
\caption{Pulse correction amplitudes and time constants for a 500~ns pulse with 200~mV amplitude and a period of $160~\mathrm{\mu s}$ and a interleaved 4000~ns pulse with -25~mV amplitude and a period of $160~\mathrm{\mu s}$. We find that some correction amplitudes scale with the pulse amplitude while others scale with both the pulse amplitude and duration (pulse area). We find that the time constants are independent of both pulse amplitude and pulse duration.}
\label{pulse_parameters}
\end{table}
\end{center}

The correction is then applied by adding
\begin{equation}
	\sum_{i=1}^{4}A_{t(b),i}\exp(-t/\tau_{t(b),i})
	\label{Corrections}
\end{equation}

to the top (bottom) segment of the pulse and adding $A_{b,\mathrm{off}}$ after the first pulse. Figs.~\ref{Pulse_corr}(d)-(f) show the shape of a corrected pulse after passing through the bias-tee. 
In principle, the corrections can be improved by repeating this process, but in practice, further improvements are limited by the vertical resolution of the AWG (0.5 mV).
Due to the frequency dependent attenuation the microwave amplitude, $A_{\varepsilon}$ at the sample changes as $f_{\mathrm{\mu w}}$ is changed. This was corrected for by measuring $A_{\varepsilon}$ at each $f_{\mathrm{\mu w}}$ used after passing a mircowave through the room temperature bias-tee and adjusting $A_{\varepsilon}$ until it was within $10\%$ of $A_{\varepsilon} = 25~\mathrm{\mu eV}$.

\section*{\hl{S6. Values of $T_2^*$ extracted from FID data}}

\begin{center}
\begin{table}[h]
\begin{tabular}{c|l|l}
	\hl{$dE/d\varepsilon$} & \hl{$T_{2,\mathrm{Gaussian}}^*$ (ns)}  & \hl{$T_{2,\mathrm{Exp}}^*$ (ns)} \\
	\hline
	\hl{$0.0399$} & \hl{$5.0 \pm 0.3$} & \hl{$5.3 \pm 0.3$} \\
	\hl{$0.0289$} & \hl{$5.1 \pm 0.2$} & \hl{$4.7 \pm 0.3$} \\
	\hl{$0.0283$} & \hl{$9.7 \pm 1.2$} & \hl{$10.1 \pm 0.7$} \\
	\hl{$0.0275$} & \hl{$5.8 \pm 0.3$} & \hl{$6.3 \pm 0.6$} \\
	\hl{$0.0265$} & \hl{$6.4 \pm 0.4$} & \hl{$6.2 \pm 0.4$} \\
	\hl{$0.0190$} & \hl{$10.4 \pm 0.8$} & \hl{$17.4 \pm 2.6$}\\
	\hl{$0.0167$} & \hl{$15.4 \pm 1.0$} & \hl{$19.2 \pm 3.0$}\\
	\hl{$0.0122$} & \hl{$22.8 \pm 1.6$} & \hl{$22.1\pm 1.1$}\\
	\hl{$0.0056$} & \hl{$73.1 \pm 7.8$} & \hl{$67.6 \pm 4.2$}\\
	\hl{$0.0042$} & \hl{$45.3 \pm 4.4$} & \hl{$66.7 \pm 5.6$}\\
	\hl{$0.0029$} & \hl{$112 \pm 10 $} & \hl{$136 \pm 24$}\\
	\hl{$0.0025$} & \hl{$177 \pm 9$} & \hl{$215 \pm 23 $}\\

\end{tabular}
\caption{\hl{Values of $T_2^*$ for fits to both a Gaussian and an exponential decay as a function of $dE/d\varepsilon$. A fit to Eq.~1 of the main text yields $\sigma_{\varepsilon} = 4.42 \pm 0.44 \mathrm{\mu eV}$ for the $T_{2,\mathrm{Exp}}^*$ values and $\sigma_{\varepsilon} = 4.39 \pm 0.32 \mathrm{\mu eV}$ for the $T_{2,\mathrm{Gaussian}}^*$ values}.}
\label{t2values}
\end{table}
\end{center}

\end{document}